\documentclass[sigconf,nonacm]{acmart}

\graphicspath{ {./img/} }
\usepackage{subcaption}
\usepackage{fontawesome5}
\usepackage{listings}
\definecolor{codegreen}{rgb}{0.06,0.57,0.11}
\definecolor{codegray}{rgb}{0.4,0.4,0.4}
\definecolor{codepurple}{rgb}{0.72,0,0.72}
\lstdefinelanguage{logs}
  {
    sensitive=false,
    morecomment=[l]{//},
    morecomment=[s]{/*}{*/},
    morestring=[b]",
  }
\lstdefinestyle{mystyle}{
    commentstyle=\color{codegreen},
    numberstyle=\color{codegray},
    keywordstyle=\color{codepurple},
    basicstyle=\ttfamily\footnotesize,
    breakatwhitespace=false,
    breaklines=true,
    captionpos=b,
    keepspaces=true,
    showspaces=false,
    showstringspaces=false,
    showtabs=false,
    tabsize=1
}
\usepackage{framed}
\usepackage{xcolor}
\colorlet{promptshade}{gray!10}
\colorlet{promptborder}{black!30}
\newenvironment{promptbox}{%
  
  \MakeFramed {
      \advance
      \hsize-\width
      \FrameRestore}%
      \raggedright 
      \ttfamily
}{
  \endMakeFramed
}
\colorlet{testimonialshade}{yellow!10}
\colorlet{testimonialborder}{black!30}
\newenvironment{testimonialbox}{%
  
  \MakeFramed {
      \advance
      \hsize-\width
      \FrameRestore}%
      \raggedright 
      \ttfamily \small
}{
  \endMakeFramed
}
\usepackage[nolist]{acronym} 
\makeatletter
\let\orig@ac\ac
\renewcommand{\ac}[1]{%
  \@ifundefined{AC@#1}{%
    \PackageError{acronym}{Undefined acronym `#1`}{%
      Define it using \string\acro{#1}{...}%
    }%
  }{%
    \orig@ac{#1}%
  }%
}
\makeatother

\begin{acronym}

\newacro{GoogleUSA}[Google]{}
\newacro{AutoDebug}[\texttt{Auto-Diagnose}]{\texttt{Auto-Diagnose}}
\newacro{TotalEvalTeams}[$39$]{$39$}
\newacro{TotalEvalExperts}[$3$]{$3$}
\newacro{TotalEvalExamples}[$71$]{$71$}
\newacro{TotalEvalExamplesGood}[$64$]{$64$}
\newacro{TotalEvalExamplesBad}[$7$]{$7$}
\newacro{TotalEvalAccuracy}[$90.14\%$]{$90.14\%$}

\newacro{AutoDebugLaunchDate}[May 2025]{May 2025}

\newacro{TotalChanges}[$91,130$]{$91,130$}
\newacro{TotalAuthors}[$22,962$]{$22,962$}
\newacro{TotalInvocations}[$224,782$]{$224,782$}
\newacro{TotalTargets}[$52,635$]{$52,635$}

\newacro{TotalAnalyzers}[$370$]{$370$}
\newacro{AutoDebugFeedbackRateRank}[$\#134$]{$\#134$}

\newacro{TotalFeedbacks}[$517$]{$517$}
\newacro{TotalFeedbacksGivers}[$437$]{$437$}
\newacro{TotalFeedbacksChanges}[$458$]{$458$}
\newacro{TotalFeedbacksPleaseFix}[$436$]{$436$}
\newacro{TotalFeedbacksPleaseFixPct}[$84.3\%$]{$84.3\%$}
\newacro{TotalFeedbacksPleaseFixReviewers}[$370$]{$370$}
\newacro{TotalFeedbacksUseful}[$51$]{$51$}
\newacro{TotalFeedbacksUsefulPct}[$9.9\%$]{$9.9\%$}
\newacro{TotalFeedbacksUsefulAuthors}[$43$]{$43$}
\newacro{TotalFeedbacksNotUseful}[$30$]{$30$}
\newacro{TotalFeedbacksNotUsefulPct}[$5.8\%$]{$5.8\%$}
\newacro{TotalFeedbacksNotUsefulAuthors}[$28$]{$28$}
\newacro{TotalFeedbacksHelpfulnessRate}[$62.96\%$]{$62.96\%$}
\newacro{AutoDebugHelpfulnessRank}[$\#14$]{$\#14$}
\newacro{AutoDebugHelpfulnessRankPct}[$3.78\%$]{$3.78\%$}
\newacro{TotalInterviewedDevelopers}[$11$]{$11$}

\newacro{FeedbackRatioAllMean}[$0.1042$]{$0.1042$}
\newacro{FeedbackRatioAllMedian}[$0.0012$]{$0.0012$}
\newacro{FeedbackRatioAutoDebug}[$0.0023$]{$0.0023$}

\newacro{FailingTestLogFilesMean}[$26$]{$26$}
\newacro{FailingTestLogFilesMedian}[$16$]{$16$}
\newacro{FailingTestLogLinesMean}[$11,058$]{$11,058$}
\newacro{FailingTestLogLinesMedian}[$2,801$]{$2,801$}
\newacro{FailingTestTokensInputMean}[$110,617$]{$110,617$}
\newacro{FailingTestTokensOutputMean}[$5,962$]{$5,962$}

\newacro{EngSatSize}[$6,059$]{$6,059$}
\newacro{DeveloperA}[\texttt{P-1}]{\texttt{P-1}}
\newacro{DeveloperB}[\texttt{P-7}]{\texttt{P-7}}
\newacro{DeveloperC}[\texttt{P-9}]{\texttt{P-9}}
\newacro{DeveloperD}[\texttt{P-3}]{\texttt{P-3}}
\newacro{DeveloperE}[\texttt{P-1}]{\texttt{P-1}}
\newacro{DeveloperF}[\texttt{P-2}]{\texttt{P-2}}
\newacro{DeveloperG}[\texttt{P-3}]{\texttt{P-3}}
\newacro{DeveloperH}[\texttt{P-4}]{\texttt{P-4}}

\end{acronym}

\begin{document}
\title{LLM-Based Automated Diagnosis Of Integration Test Failures At Google}

\author{Celal Ziftci}
\email{celal@google.com}
\affiliation{
  \institution{\acs{GoogleUSA}}
  \city{New York}
  \state{NY}
  \country{USA}
}
\author{Ray Liu}
\email{rayliu@google.com}
\affiliation{
  \institution{\acs{GoogleUSA}}
  \city{New York}
  \state{NY}
  \country{USA}
}
\author{Spencer Greene}
\email{greenespencer@google.com}
\affiliation{
  \institution{\acs{GoogleUSA}}
  \city{New York}
  \state{NY}
  \country{USA}
}
\author{Livio Dalloro}
\email{dalloro@google.com}
\affiliation{
  \institution{\acs{GoogleUSA}}
  \city{New York}
  \state{NY}
  \country{USA}
}

\begin{abstract}

Integration testing is critical for the quality and reliability of complex software systems. However, diagnosing their failures presents significant challenges due to the massive volume, unstructured nature, and heterogeneity of logs they generate. These result in a high cognitive load, low signal-to-noise ratio, and make diagnosis difficult and time-consuming. Developers complain about these difficulties consistently and report spending substantially more time diagnosing integration test failures compared to unit test failures.

To address these shortcomings, we introduce \acs{AutoDebug}, a novel diagnosis tool that leverages Large Language Models (LLMs) to help developers efficiently determine the root cause of integration test failures. \acs{AutoDebug} analyzes failure logs, produces concise summaries with the most relevant log lines, and is integrated into Critique, Google's internal code review system, providing contextual and in-time assistance.

Based on our case studies, \acs{AutoDebug} is highly effective. A manual evaluation conducted on \acs{TotalEvalExamples} real-world failures demonstrated \acs{TotalEvalAccuracy} accuracy in diagnosing the root cause. Following its Google-wide deployment, \acs{AutoDebug} was used across \acs{TotalTargets} distinct failing tests. User feedback indicated that the tool was deemed "Not helpful" in only \acs{TotalFeedbacksNotUsefulPct} of cases, and it was ranked \acs{AutoDebugHelpfulnessRank} (in the top \acs{AutoDebugHelpfulnessRankPct}) in helpfulness among \acs{TotalAnalyzers} tools that post findings in Critique. Finally, user interviews confirmed the perceived usefulness of \acs{AutoDebug} and positive reception of integrating automatic diagnostic assistance into existing workflows.

We conclude that LLMs are highly successful in diagnosing integration test failures due to their capacity to process and summarize complex textual data. Integrating such AI-powered tooling automatically into developers’ daily workflows is perceived positively, with the tool's accuracy remaining a critical factor in shaping developer perception and adoption.

\end{abstract}

\ccsdesc[500]{Software and its engineering~Software testing and debugging}
\ccsdesc[500]{Computing methodologies~Information extraction}
\ccsdesc[300]{Software and its engineering~Empirical software validation}
\ccsdesc[100]{Software and its engineering~Maintaining software}

\keywords{Software, Testing, Debugging, Diagnosis, Productivity, LLM}

\maketitle
\section*{Preprint Notice}
This is a preprint of a paper accepted at the IEEE/ACM 48th International Conference on Software Engineering (ICSE) 2026.

\section{Introduction}

Software testing is a crucial phase in the software development lifecycle, ensuring the quality and reliability of complex systems. Unit testing and integration testing are two of the most commonly employed types of software testing \citep{pressman2014software}.

Unit testing focuses on individual components or modules of a software application. The primary goal of unit tests is to verify that each unit of the software performs as expected in isolation. Developers typically write these tests during the coding phase, and they are often automated. They help in early detection of bugs, facilitate easier debugging, and provide documentation for the individual components.

Integration testing verifies the interaction, communication, and data exchange between separate software components that are typically already unit-tested \citep{humble2010continuous,pressman2014software}. In modern, complex, distributed software systems, integration testing is a critical part of quality assurance. It aims to expose defects in the interfaces and interactions between integrated components, rather than within the components themselves. This type of testing is essential for ensuring that different parts of a system work together seamlessly to achieve the desired functionalities.

When these tests fail, the primary artifacts available for post-mortem analysis and debugging are the semi-structured text logs \citep{he2017log,he2021survey} generated by the test runtime itself and the system under test (SUT). However, as software systems have scaled, the manual analysis of these logs has become infeasible. Developers face significant challenges, including massive log volumes \citep{he2017log}, unstructured and heterogeneous text formats \citep{yuan2012characterizing}, and a low signal-to-noise ratio where crucial errors are often obscured by routine operational data \citep{zhao2023log}. This complexity results in a high cognitive load required to interpret the failure context across numerous log files.

The difficulty of this process is empirically documented within Google: diagnosing integration test failures was identified as one of the top five most frequent complaints in a company-wide survey \citep{GoogleEngSat} of \acs{EngSatSize} developers. Furthermore, developers consistently report spending substantially more time diagnosing integration test failures, often more than an hour and sometimes exceeding a day, compared to unit test failures, highlighting the inherent difficulty of this task. The scale of the challenge is reflected in the properties of failing tests analyzed, which had a median of \acs{FailingTestLogFilesMedian} log files and \acs{FailingTestLogLinesMedian} log lines.

While many automated program diagnosis and repair techniques exist, from statistical debugging to spectrum-based analysis, recent work has increasingly explored the capabilities of Large Language Models (LLMs). However, this research, along with its associated benchmarks, has concentrated primarily on small-scale or unit-test-level debugging and repair. The unique challenges of integration testing, involving multiple interacting components, complex environment setups, and massive, distributed log outputs, remain largely under-explored.

To address the shortcomings and inefficiency of manual log investigation, this paper proposes \acs{AutoDebug}, a novel failure diagnosis tool leveraging LLMs. Given the proficiency of LLMs in text processing and summarization tasks, \acs{AutoDebug} is designed to help developers efficiently determine the root cause of complex integration test failures. This functionality is integrated into Google’s internal code review system Critique \citep{critique}, providing contextual and in-time assistance to streamline the debugging process.

The effectiveness of \acs{AutoDebug} was evaluated through quantitative and qualitative methods. A manual evaluation on \acs{TotalEvalExamples} real-world integration test failures demonstrated a high accuracy rate of \acs{TotalEvalAccuracy} in detecting and summarizing the root cause. Following its deployment, \acs{AutoDebug} was used across \acs{TotalTargets} distinct tests. User feedback gathered from code changes indicated that the tool was deemed "Not helpful" in only \acs{TotalFeedbacksNotUsefulPct} of cases. User interviews confirmed the perceived usefulness and positive reception of integrating automatic diagnostic assistance into existing workflows.

We conclude that LLMs are highly successful in diagnosing integration test failures due to their capacity to process and summarize complex textual data, and developers prefer using such tooling when integrated into their workflows.

\section{Diagnosing Integration Test Failures}

In this section, we discuss our motivation to focus specifically on integration tests, showcase a specific type of integration test at Google with its various characteristics on a motivating example, explain how code reviews are conducted and integration test failures are reported during the code review process, showcase how developers diagnose test failures traditionally, and finally discuss the shortcomings of manual diagnosis.

\subsection{Why Focus On Integration Tests?}

Our work focuses on diagnosing integration tests rather than unit tests, informed by several internal surveys conducted at Google.

First, in a Google-wide survey of \acs{EngSatSize} developers about their use of and experience on various internal developer tools, named \texttt{EngSat} \citep{GoogleEngSat}, diagnosing integration test failures was identified as one of the top five most frequent complaints.

\begin{figure}[tbp]
    \centering
    
    \begin{minipage}{\linewidth}
        \centering
        \includegraphics[width=\linewidth]{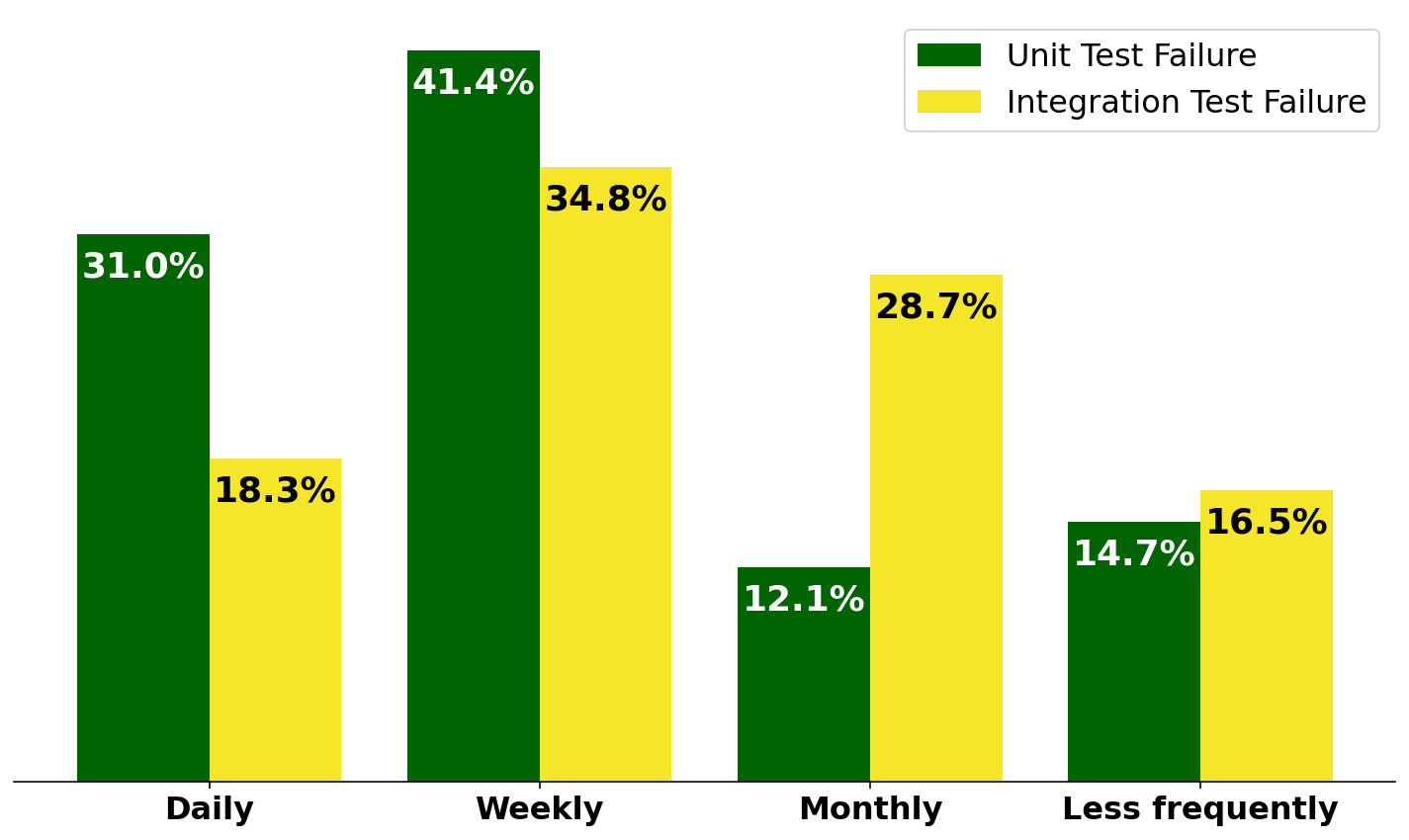}
        \subcaption{Frequency of encountering unit and integration test failures}\label{fig:unit-vs-integration-encounter}
    \end{minipage}

    \vspace{0.3cm}

    \begin{minipage}{\linewidth}
        \centering
        \includegraphics[width=\linewidth]{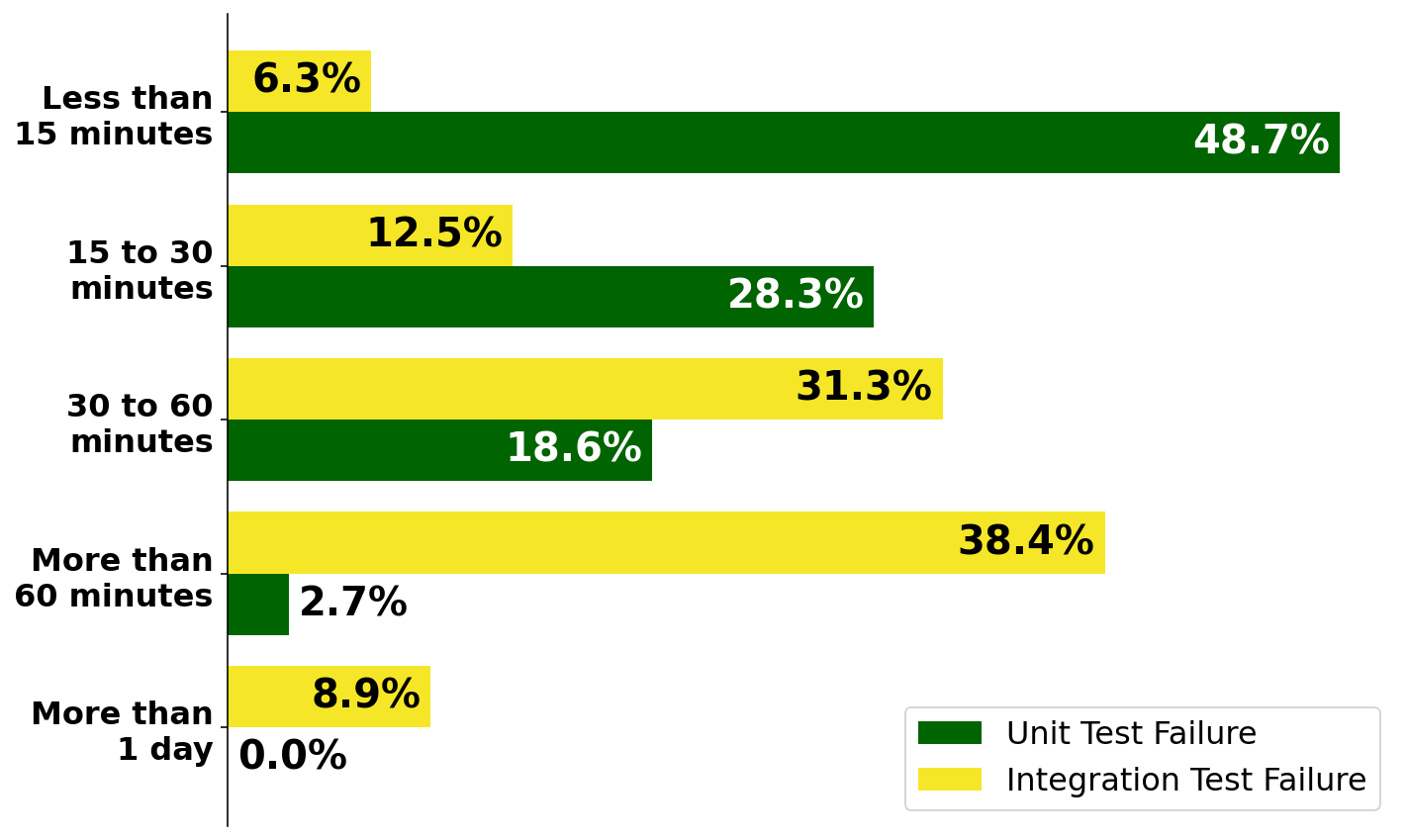}
        \subcaption{Duration of failure diagnosis of unit and integration test failures}\label{fig:unit-vs-integration-diagnosis}
    \end{minipage}
    \Description{Integration test encounter and diagnosis breakdowns}
    \caption{Differences in the encounter and diagnosis of unit and integration test failures as reported by 116 survey respondents.}
    \label{fig:llm-hallucination-examples}
\end{figure}

To understand the differences between unit and integration test failures, we conducted a follow-up survey to ask 116 developers to self-report usage and pain points on diagnosing unit and integration tests, named \textit{Survey-2}. As shown in Figure \ref{fig:unit-vs-integration-encounter}, developers reported encountering unit test failures more often than integration test failures daily and weekly, while integration test failures are reported to occur more frequently monthly. Unit tests are typically executed earlier and more often in the software development lifecycle, while integration tests are executed later, as they are typically more expensive and slower. As a result, the survey responses support our expectations of the frequency of unit and integration test failures.

Additionally, as shown in Figure \ref{fig:unit-vs-integration-diagnosis}, developers reported spending significantly more time diagnosing integration test failures, often more than an hour and sometimes longer than a day, pointing to its difficulty and supporting the complaints identified in \textit{EngSat}. We discuss potential reasons for this in Section \ref{subsection:Diagnosing-Integration-Test-Failures}.

\subsection{Hermetic Functional Integration Tests}

At Google, there are different kinds of integration tests employed by teams, including functional, performance, reliability and security. The tests under consideration for this work are the hermetic functional tests, those that are brought up entirely within isolated environments without relying on external services or shared infrastructure, and that exercise business logic, as opposed to qualitative aspects of the system such as performance, security and reliability.

We focus on hermetic tests, as they provide a consistent and reproducible testing environment, simplifying debugging by eliminating external dependencies as potential sources of failure.

\begin{figure}
\centering
\includegraphics[scale=0.35]{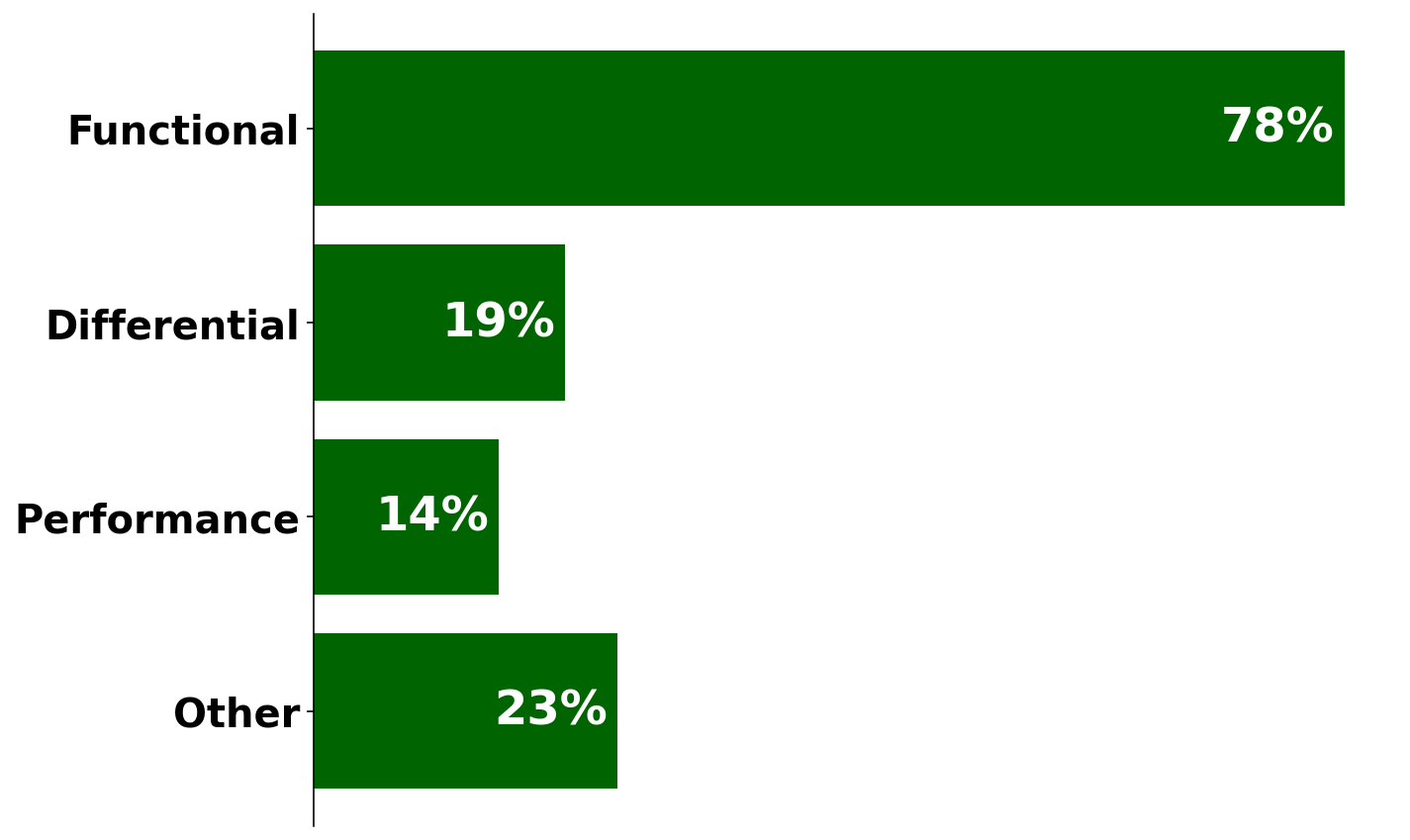}
\caption{Different kinds of integration tests employed by teams at Google, based on a survey of 239 respondents. Note that a single team can have several types of these integration tests, therefore, the sum exceeds $100\%$.}
\Description{Integration test breakdown}
\label{fig:integration-test-breakdown}
\end{figure}

We focus on functional tests, as they are the most frequently used integration tests across Google based on a survey of 239 respondents, as shown in Figure \ref{fig:integration-test-breakdown}. Note that a single team can have several types of these integration tests, therefore, the sum exceeds $100\%$.

\subsection{Integration Test Failure Findings During Code Review}

Google's internal code review system, Critique \citep{sadowski2018modern}, is a web-based interface that facilitates the code review process, enabling discussions between authors and reviewers, and ensuring adherence to code quality standards and policies.

Beyond human reviews, Critique also integrates automated analysis results crucial for maintaining code health. A key mechanism for surfacing automated analysis feedback within Critique is the \textit{findings} feature \citep{sadowski2015tricorder} which allows various tools and services to post structured annotations as comments, directly within the context of a pending change. These findings may range from style linting errors, to static analysis warnings and the results of continuous integration systems and test executions \citep{monorepo-2}.

\begin{figure}
\centering
\includegraphics[scale=0.16]{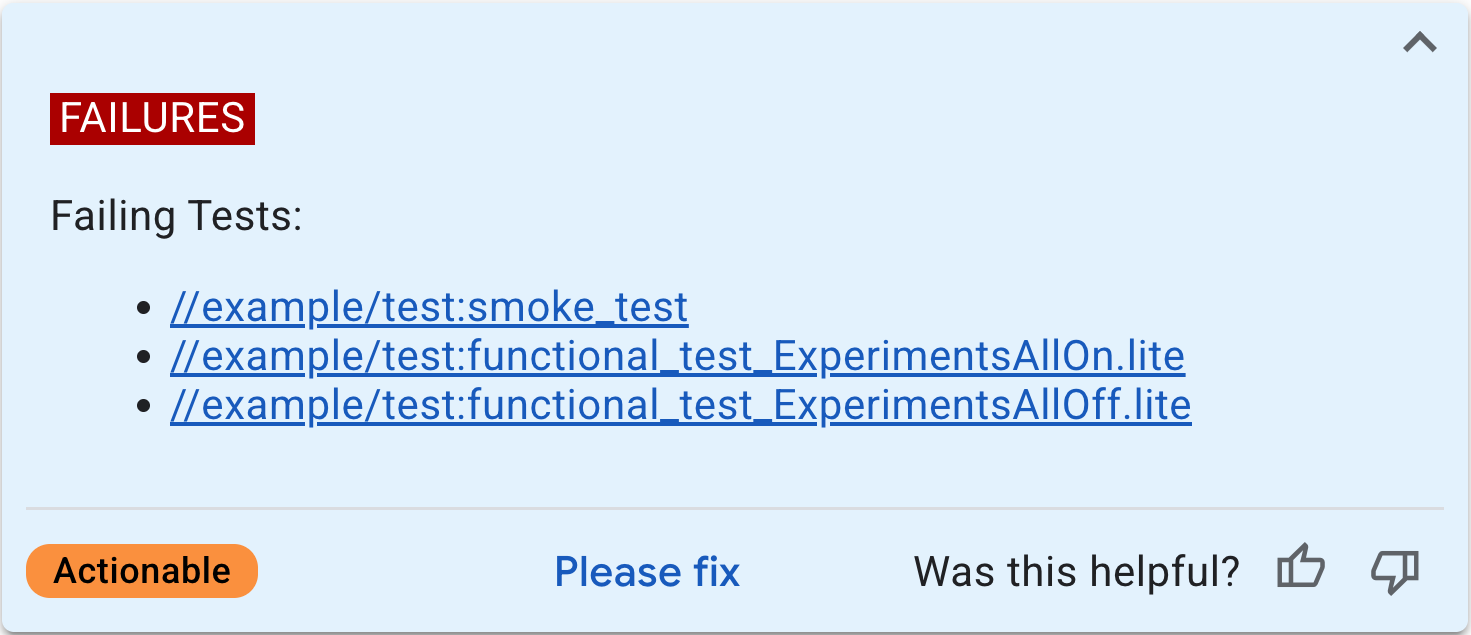}
\Description{Critique finding for failed tests}
\caption{Integration test failure information surfaced in Critique \citep{sadowski2018modern}, Google's web-based internal code review system.}
\label{fig:critique-test-failed}
\end{figure}

Relevant to test-driven development, as shown in Figure \ref{fig:critique-test-failed}, the functional hermetic integration tests discussed earlier are typically executed automatically when a code change is sent to a reviewer, and failures are surfaced as actionable findings within Critique, preventing submission until resolved. This tight integration ensures that test results are highly visible and directly linked to the code changes under review, reinforcing the practice of submitting well-tested code.

\subsection{Diagnosing Integration Test Failures Is Hard}
\label{subsection:Diagnosing-Integration-Test-Failures}

\begin{figure}
\centering
\includegraphics[scale=0.62]{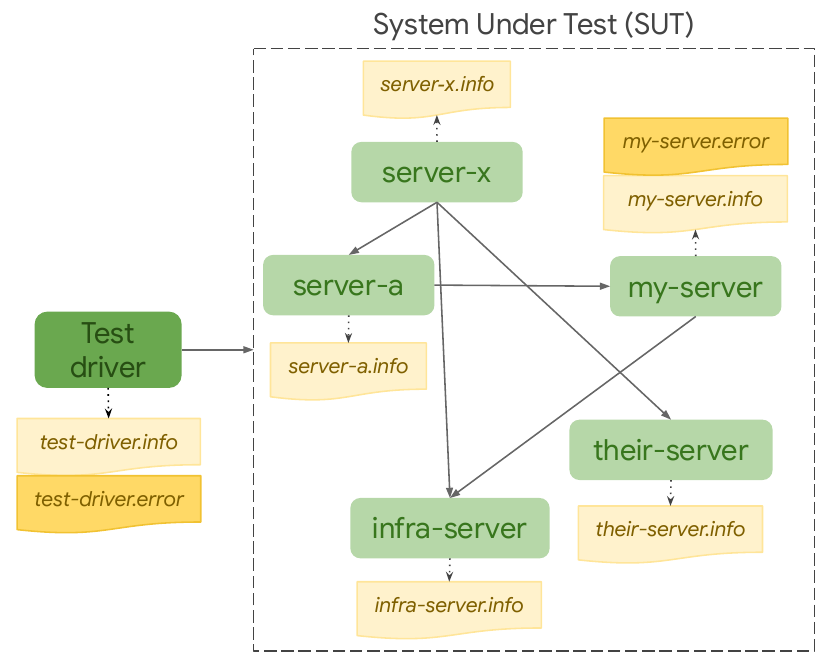}
\Description{Test driver and system under test}
\caption{Test driver and the system under test (SUT) that it controls to exercise business logic in the integration test. Each component produces its own log files, typically split by log level, and named dynamically after the component.}
\label{fig:test-driver-and-sut}
\end{figure}

Functional integration tests, as shown in Figure \ref{fig:test-driver-and-sut}, typically comprise of the test driver and the system under test (SUT). The SUT typically consists of a collection of servers, called \textit{components}, that communicate with each other. The test driver is responsible for configuring the SUT and bringing up the components needed by the test, and is itself a separate component written in languages such as C++, Java or Python.

During the execution of an integration test, logs \citep{yuan2012characterizing} are generated from several sources:
\\

\noindent \textbf{Test driver logs}: The test driver produces its own logs to record its actions, configurations, and any issues it encounters while setting up the components or interacting with the SUT.
\\

\noindent \textbf{SUT component logs}: Each individual component within the SUT generates its own logs output to distinct files, detailing its internal operations, communication with other components, and any errors or events that occur during its execution. Each component produces log files dynamically named after the component name and can produce several log files split by log level.
\\

These logs are crucial to understand the behavior of the distributed system. Both the test driver logs and the SUT component logs are collected and made available to developers for diagnosis when a test fails. The default logs shown to the developer as they investigate the failure are the test driver logs, as these are the highest level logs about the execution of the test.

When a developer receives a finding for a failing integration test on their code change during code review, shown in Figure \ref{fig:critique-test-failed}, the diagnostic process typically begins with an examination of the test driver's logs that provide a high-level summary of the test execution. However, there are several shortcomings of test driver logs, as they lack the necessary detail to pinpoint the root cause of the failure.

First, most of the time, a failure manifests as a generic error, e.g. a timeout, in the test driver logs, indicating that a component of the SUT failed to become operational within the expected time frame, or that a specific test assertion caused a failure. In these cases, as shown in Figure \ref{fig:test-driver-and-sut}, developers need to investigate the list of created log files one by one to identify which one contains the root cause.

Second, the sheer volume of logs, originating from both the test driver and numerous SUT components, presents a significant challenge \citep{he2017log}. Developers must manually sift through a multitude of log files, each with its own formatting and conventions, to find the relevant error messages.

Third, many logs contain warnings and errors that are not related to the current failure, or the execution of the test at all, as they happen to be logged at warning or error levels even though they are recoverable. This creates a low signal-to-noise ratio that can mislead the developers during investigation \citep{zhao2023log}. The cognitive load required to correlate events across different logs and distinguish between benign errors and the actual cause of the failure is substantial.

This complexity often overwhelms developers, leading them to seek alternative ways for diagnosis of the failure. Two of the most common ways are seeking assistance from colleagues with more experience in the specific system or test, and consulting the infrastructure team that builds and maintains the testing framework itself. This reliance on manual expertise results in inefficiency and scalability issues inherent in diagnosing integration test failures, highlighting the need for more efficient tools to investigate such failures.
\section{\acs{AutoDebug}: Automatically Diagnosing Integration Test Failures}

\begin{figure*}
\centering
\includegraphics[scale=0.87]{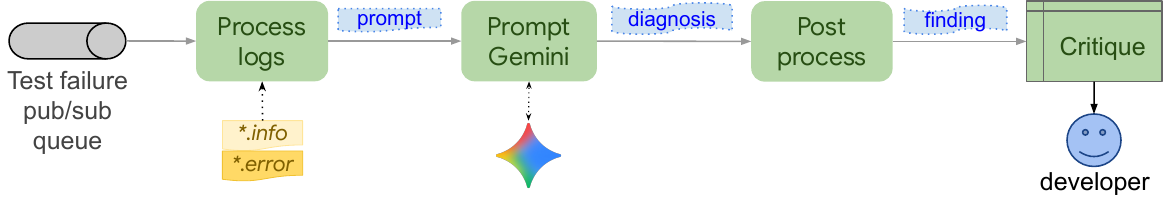}
\Description{\AutoDebug{} design}
\caption{System overview of automatically generating findings for integration test failure diagnosis with \acs{AutoDebug}.}
\label{fig:auto-debug-design}
\end{figure*}

As LLMs are known to work well on text processing and summarization tasks, we employed them on the task of processing integration test failure logs and producing a diagnosis summary for developers. In this section, we discuss the elements of the system, named \acs{AutoDebug}, shown in Figure \ref{fig:auto-debug-design}.

Upon a test failure, \acs{AutoDebug} gets notified and runs automatically, and test driver and SUT component logs at levels \texttt{INFO} and higher, spread across different data centers, processes, threads and logging levels, are joined and sorted by timestamp into a single log stream to be passed to the LLM. Listing \ref{lst:log-line} shows examples of log lines collected from the \texttt{server-a} component.

\begin{lstlisting}[label={lst:log-line}, float, language=logs, caption=Example log lines from the \texttt{server-a} component.]
server-a.info:
    2025-09-17-14:12:32 | dc7 | p41 | t-2 | file.py:444 | Server is starting

server-a.error:
    2025-09-17-16:59:41 | dc3 | p13 | t-7 | file2.py:41 | Server encountered an error, shutting down
\end{lstlisting}

Then, \acs{AutoDebug} constructs an LLM prompt to be sent to Gemini \citep{team2023gemini,comanici2025gemini} using the prompt template, shown in Figure \ref{fig:gemini-prompt}, concatenated with logs following the \texttt{<LOGS=>} section, and component metadata under the \texttt{<CONTEXT=>} section. The prompt is developed over several iterations of observing the LLM's outputs on real-world failures to include guided, step-by-step reasoning with strict negative constraints and precise output formatting, to avoid speculative, incomplete, or irrelevant diagnoses. The LLM parameters used are:

\begin{itemize}
    \item LLM = Gemini 2.5 Flash \citep{team2023gemini,comanici2025gemini} so that \acs{AutoDebug} is fast and cost effective.
    \item $temperature = 0.1$ so that LLM responses are mostly deterministic for easy debugging.
    \item $top_p = 0.8$ so that there is room for creativity while keeping out extremely unlikely tokens.
\end{itemize}

\noindent Gemini has not been trained or fine-tuned with the specific integration test failure logs that \acs{AutoDebug} ran on.

\begin{figure}
\centering
\includegraphics[scale=0.18]{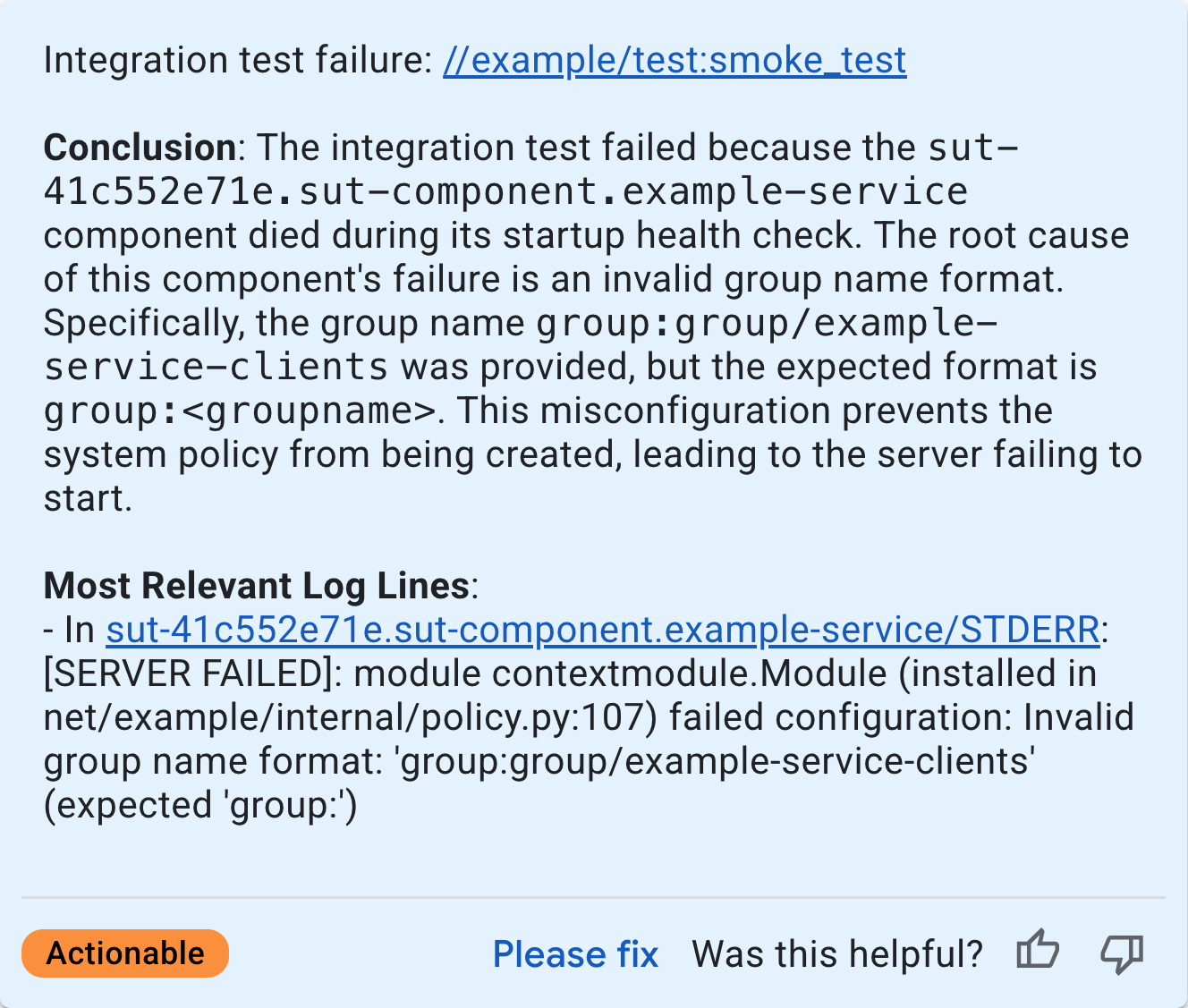}
\Description{\AutoDebug{} finding posted to Critique}
\caption{The LLM-based diagnosis result posted as a finding to Critique.}
\label{fig:auto-debug-finding}
\end{figure}

After getting the diagnosis from the LLM, \acs{AutoDebug} post-processes its response to show the output in markdown format, to convert the log lines it provides to links, and to produce a finding to show in Critique, as shown in Figure \ref{fig:auto-debug-finding}. Developers may find this diagnosis useful as is, or may click the links to investigate the failure further.

\begin{figure*}[htbp]
\centering
\begin{promptbox}
You are helping developers at Google understand the root cause of a failed integration test. Available log lines about the test are listed under <LOGS=>, and context about SUT components are listed under <CONTEXT=>.\\[0.5em]

Your goal is to find the root cause from the log lines under <LOGS=> and potentially <CONTEXT=>. Let's think about the analysis step-by-step and show your thought process after each step:\\
1. Scan all sections in <LOGS=> as subsection headers. No need to print all of them out at this point, but you will need to refer to them later.\\
2. If <CONTEXT=> is provided, read it and treat it as context and potentially instructions to help you understand the log lines and debug test failures better.\\
3. Inspect the section that contains test failures.\\
4. Inspect the logs section if it exists. Summarize the errors, and what component failed.\\
5. Inspect the other sections and summarize what those lines are signaling and print the most likely cause of the error. Also take into consideration of the command line arguments passed to the component, and see if the error is related to the arguments. For example, if the command line argument contains a keyword that's also mentioned in the error message, it's likely that the argument is a contributing factor to the error.\\
6. Try to reach a conclusion on the root cause of the failure. You must not skip the rest of the steps.\\
7. Judge if you have enough information to reach a conclusion in step 6. You must ruthlessly adhere to the following rules:\\
  - If the logs clearly point out the SUT component that failed to start up, you MUST locate the corresponding component log lines and *ONLY* use those to reach a conclusion.\\
  - If the logs do not contain any log lines from the component that failed to become healthy, you *MUST NOT* draw any conclusion from the information you have.\\
  - You MUST not make any assumptions about the SUT infrastructure: if specific SUT components are pointed out, you MUST *ONLY* use those components' log lines to reach a conclusion.\\
  - Any conclusion about a SUT component must be based on its log lines, and you MUST NOT draw any conclusions by guessing.\\
8. From step 7, if you don't understand the root cause of the failure, think about what other information you might need. For example, if there are errors complaining about processes not mentioned in the log lines, you should mention in your response that you need access to those logs and you must not draw any conclusion from the information you have.\\[0.5em]

Things to keep in mind:\\
 - Every test will have a log about the test exiting due to SIGINT. This is normal and not the cause of the failure.\\
 - The response should be in human readable sentences. The format depends on the conclusion of your investigation (i.e. if step 7 is satisfied and if step 8 is performed).\\[0.5em]

If you reached a conclusion in step 6 and verified that you have enough information to reach that conclusion in step 7, start with the conclusion of your investigation under a "==Conclusion==" header. This is the most important part of the response.\\[0.5em]

Finally, list the steps you have taken in the process under a "==Investigation Steps==" header. After that, print the most relevant log lines under a "==Most Relevant Log Lines==" header. These log lines must be from the sections printed in step-5. For each log line, print it in the following format:\\
- log-file-name: <log-file-name> (e.g. foo-bar.info, must be the same as the log-file-name in the section header)\\
- timestamp: <timestamp-in-the-format-of-YYYY-MM-DD-hh:mm:ss> (e.g. 2025-03-27-06:00:000, if you can't find the timestamp, leave this empty)\\
- callsite: <callsite-file-name:callsite-line> (e.g. foo/bar/baz.cc:123, if callsite-filename is long, only include the suffix)
**content**: <root-cause-relevant-part-of-the-log-line> (e.g. The server encountered an error: the root cause of the failure)\\[0.5em]

This log line format is important and should be followed exactly. It's VERY IMPORTANT to only include the interesting part of the log line content.\\[0.5em]

<LOGS=>\\
\%s
\\[0.5em]
<CONTEXT=>\\
\%s
\end{promptbox}

\Description{Gemini prompt}
\caption{The prompt template used to construct the prompt sent to the LLM.}
\label{fig:gemini-prompt}
\end{figure*}

\section{Evaluation}

In this section, we discuss how we evaluate \acs{AutoDebug}, describe our case studies, and report the results of our interviews on the usability of our system.

\subsection{Manual Evaluation}
\label{section:manual-evaluation}

To evaluate the effectiveness of \acs{AutoDebug}, we first conducted a case study where we ran \acs{AutoDebug} on \acs{TotalEvalExamples} randomly selected integration test failures from \acs{TotalEvalTeams} distinct teams across Google, summarized in Table \ref{table:manual-eval-stats}. We asked \acs{TotalEvalExperts} expert developers with at least five years of experience each from the integration test infrastructure team to assess the diagnoses, similar to the one in Figure \ref{fig:auto-debug-finding}, and report whether the \texttt{Conclusion} or any of the reported \texttt{Most Relevant Log Lines} provide accurate context for the root cause of the failure. We then held a meeting to go over their reports and align on any disagreements, to obtain final agreement on the assessment for each failure. Based on the experts' reporting, \acs{AutoDebug} was successful in diagnosing the root cause on \acs{TotalEvalExamplesGood} of the failures, resulting in a \acs{TotalEvalAccuracy} success rate.

\begin{table}
\centering
\caption{Aggregate statistics about the manual evaluations.}
\begin{tabular}{r|l}
\# teams that owned the failures evaluated & 39 \\
\# developers that conducted evaluations & 3 \\
\# failures evaluated & \acs{TotalEvalExamples} \\
\# accurate diagnoses on the failures evaluated & \acs{TotalEvalExamplesGood} (\acs{TotalEvalAccuracy}) \\
\end{tabular}
\label{table:manual-eval-stats}
\end{table}

We also asked the three expert developers to investigate where \acs{AutoDebug} may have failed to diagnose the root cause on the remaining \acs{TotalEvalExamplesBad} failures. Based on deeper investigation, two distinct issues were discovered, in 4 of the cases, test driver log files were not properly saved when it crashed, and in 3 of the cases, SUT component log files were not properly saved when the component crashed, both bugs discovered in the test infrastructure and reported to the relevant teams.

\subsection{Production Usage and User Feedback}
\label{section:user-feedback}

After the manual evaluation, we launched \acs{AutoDebug} to automatically run on all integration test  failures during code changes across Google's code repository \citep{monorepo} starting \acs{AutoDebugLaunchDate}, with detailed statistics listed in Table \ref{table:usage-stats}.

We showed Critique findings, as shown in Figure \ref{fig:auto-debug-finding}, on \acs{TotalChanges} code changes from \acs{TotalAuthors} distinct authors for \acs{TotalInvocations} executions of \acs{TotalTargets} distinct tests.

For test failures, it is important to post any diagnostic help findings to Critique quickly so that developers do not change context, or start manually debugging the failure themselves. \acs{AutoDebug} took 56 seconds in the median (p50), and 346 seconds in the 90th percentile (p90) to post its findings to Critique, significantly faster than the time developers typically spend on debugging integration tests, discussed in Figure \ref{fig:unit-vs-integration-diagnosis}.

The failing integration tests had a mean of \acs{FailingTestLogFilesMean} log files and a median of \acs{FailingTestLogFilesMedian} log files, and a mean of \acs{FailingTestLogLinesMean} log lines and a median of \acs{FailingTestLogLinesMedian} log lines, in line with the difficulties of diagnosing integration test failures discussed in Section \ref{subsection:Diagnosing-Integration-Test-Failures}. Executions of \acs{AutoDebug} had a mean of \acs{FailingTestTokensInputMean} input tokens and \acs{FailingTestTokensOutputMean} output tokens, making it highly cost effective.

\begin{table}
\centering
\caption{Aggregate statistics about \acs{AutoDebug} findings posted to Critique and the properties of the failing integration tests it diagnosed.}
\begin{tabular}{r|l}
\# total code changes & \acs{TotalChanges} \\
\# total authors of the code changes & \acs{TotalAuthors} \\
\# distinct tests analyzed by \acs{AutoDebug} & \acs{TotalTargets} \\
\# total executions of \acs{AutoDebug} & \acs{TotalInvocations} \\
\hline
p50 time to post \acs{AutoDebug} finding to Critique & 56 sec \\
p90 time to post \acs{AutoDebug} finding to Critique & 346 sec \\
\hline
Mean \# log files per failing test & \acs{FailingTestLogFilesMean} \\
Median \# log files per failing test & \acs{FailingTestLogFilesMedian} \\
Mean \# log lines per failing test & \acs{FailingTestLogLinesMean} \\
Median \# log lines per failing test & \acs{FailingTestLogLinesMedian} \\
\hline
Mean \# input tokens per \acs{AutoDebug} execution & \acs{FailingTestTokensInputMean} \\
Mean \# output tokens per \acs{AutoDebug} execution & \acs{FailingTestTokensOutputMean}
\end{tabular}
\label{table:usage-stats}
\end{table}

On code changes, as shown in the bottom section of Figure \ref{fig:auto-debug-finding}, developers can click on several buttons to interact with the findings:

\begin{itemize}
    \item \textbf{Please fix -- $\mathbf{PF}$}: A reviewer can click this to ask the author of the code change to fix it. Note that the same "Please fix" button exists on the test failure finding itself, as shown in Figure \ref{fig:critique-test-failed}. Typically, a reviewer would click on "Please fix" on that finding if they want the test to be fixed, while they would click "Please fix" on the \acs{AutoDebug} finding if they intend to ask the author to make use of the diagnosis.
    \item \textbf{\faThumbsUp{} Helpful -- $\mathbf{H}$}: An author can click this to give feedback to the tool owner that the diagnosis was helpful.
    \item \textbf{\faThumbsDown{} Not helpful -- $\mathbf{N}$}: An author can click this to give feedback to the tool owner that the diagnosis was not helpful.
\end{itemize}

To better understand user engagement and feedback, we compared \acs{AutoDebug} with other tools that post findings in Critique across Google on several dimensions.
\\

\begin{table}
\centering
\caption{Statistics about the feedback posted to Critique by different tools and \acs{AutoDebug}.}
\begin{tabular}{r|l}
\# tools that post findings to Critique & \acs{TotalAnalyzers} \\
\hline
Mean feedback-rate across all findings & \acs{FeedbackRatioAllMean} \\
Median feedback-rate across all findings & \acs{FeedbackRatioAllMedian} \\
\hline
Feedback-rate for \acs{AutoDebug} findings & \acs{FeedbackRatioAutoDebug} \\
Feedback-rate rank of \acs{AutoDebug} among all tools & \acs{AutoDebugFeedbackRateRank}
\end{tabular}
\label{table:feedback-rate-stats}
\end{table}

\noindent \textbf{\texttt{Feedback-rate}}: This metric measures user engagement, i.e. how often developers report feedback upon a finding on a code change in Critique. Listed in Table \ref{table:feedback-rate-stats}, among \acs{TotalAnalyzers} tools that posted at least 100 findings to Critique, \acs{AutoDebug} ranked \acs{AutoDebugFeedbackRateRank} with a feedback-rate of \acs{FeedbackRatioAutoDebug}, compared to a median of \acs{FeedbackRatioAllMedian} and a mean of \acs{FeedbackRatioAllMean} across all findings. Developers were engaged with \acs{AutoDebug} in line with other tool findings posted to their code changes.
\\

\begin{table}
\centering
\caption{Aggregate statistics about the user feedback for \acs{AutoDebug}.}
\begin{tabular}{r|l}
\# total feedback & \acs{TotalFeedbacks} \\
\# total code changes with feedback & \acs{TotalFeedbacksChanges} \\
\# distinct developers that reported feedback & \acs{TotalFeedbacksGivers} \\
\hline
\underline{Please fix -- PF} & \\
\# total "Please fix" & \acs{TotalFeedbacksPleaseFix} (\acs{TotalFeedbacksPleaseFixPct}) \\
\# distinct developers that clicked "Please fix" & \acs{TotalFeedbacksPleaseFixReviewers} \\
\hline
\underline{\faThumbsUp{} Helpful -- H} & \\
\# total "Helpful" & \acs{TotalFeedbacksUseful} (\acs{TotalFeedbacksUsefulPct}) \\
\# distinct developers that clicked "Helpful" & \acs{TotalFeedbacksUsefulAuthors} \\
Helpfulness-rate ($\frac{H}{H + N}$) & \acs{TotalFeedbacksHelpfulnessRate} \\
Helpfulness-rate rank across all tools & \acs{AutoDebugHelpfulnessRank} out of \acs{TotalAnalyzers} \\

\hline
\underline{\faThumbsDown{} Not Helpful -- N} & \\
\# total "Not helpful" & \acs{TotalFeedbacksNotUseful} (\acs{TotalFeedbacksNotUsefulPct}) \\
\# distinct developers that clicked "Not Helpful" & \acs{TotalFeedbacksNotUsefulAuthors} \\
Not-helpful-rate ($\frac{N}{PF + H + N}$) & \acs{TotalFeedbacksNotUsefulPct}
\end{tabular}
\label{table:user-feedback}
\end{table}

Table \ref{table:user-feedback} lists statistics of feedback reports we received from developers: a total of \acs{TotalFeedbacks} feedback were reported by \acs{TotalFeedbacksGivers} distinct users, \acs{TotalFeedbacksPleaseFix} of these reports were "Please fix" reported by \acs{TotalFeedbacksPleaseFixReviewers} reviewers, \acs{TotalFeedbacksUseful} of the reports were "Helpful" reported by \acs{TotalFeedbacksUsefulAuthors} distinct authors, and \acs{TotalFeedbacksNotUseful} of the reports were "Not helpful" reported by \acs{TotalFeedbacksNotUsefulAuthors} distinct authors.
\\

\noindent \textbf{\texttt{Not-helpful-rate}}: This metric measures $\frac{N}{PF + H + N}$, the percentage of "Not helpful" feedback reports across all reports. To avoid noisy tools and annoying developers with bad findings, for a tool to continue posting findings in Critique, its not-helpful-rate has to stay below $10\%$ as a general guideline \citep{sadowski2015tricorder}. \acs{AutoDebug} has a \acs{TotalFeedbacksNotUsefulPct} not-helpful-rate, in line with the accuracy results of our manual evaluation.
\\

\noindent \textbf{\texttt{Helpfulness-rate}}: This metric measures $\frac{H}{H + N}$, the ratio of "Helpful" feedback reports to all code change author feedback reports. This ratio is important as it reflects that the diagnoses produce good findings. For \acs{AutoDebug}, this metric is currently \acs{TotalFeedbacksHelpfulnessRate}, with more "Useful" feedback than "Not useful".

The reported helpfulness-rate of \acs{TotalFeedbacksHelpfulnessRate} is lower compared to our manual evaluation with \acs{TotalEvalAccuracy} accurate diagnoses. It is a well-established phenomenon in psychology that users are more motivated to report negative experiences than positive ones \citep{rozin2001negativity}. Therefore, in feedback mechanisms such as thumbs up/down buttons, common in modern developer tools \citep{shi2025natural}, it is expected that negative feedback will be disproportionately represented, as positive or 'as-expected' interactions often go unreported. We propose that this phenomenon plays a role in the \acs{AutoDebug} user feedback.

Furthermore, many of the tools that post findings to Critique do not use AI or LLMs, they are deterministic checks on various aspects of code such as lints, tests and style checks. Across \acs{TotalAnalyzers} different types of tools that post findings to Critique, at the time of writing, \acs{AutoDebug} ranks as \acs{AutoDebugHelpfulnessRank} (in the top \acs{AutoDebugHelpfulnessRankPct}).

Given the high rank, we propose that developers have been frustrated with diagnosing integration test failures as identified with surveys, and \acs{AutoDebug} diagnoses resonate with them when they are helpful.

\subsection{User interviews}

We conducted \acs{TotalInterviewedDevelopers} in-person interviews with code change authors that reported "Useful" and "Not useful", and code change reviewers that reported "Please fix" feedback, with excerpts in Figure \ref{fig:testimonials}. In this section, we summarize key learnings based on the participants' responses.

First, several participants (e.g. \acs{DeveloperE}, \acs{DeveloperF}, \acs{DeveloperG}) discussed the difficulty of diagnosing failing integration tests, with reasons similar to those discussed in Section \ref{subsection:Diagnosing-Integration-Test-Failures}.

Several participants (e.g. \acs{DeveloperH}, \acs{DeveloperA}, \acs{DeveloperB}) reported that they found the diagnoses useful. \acs{DeveloperA} highlighted that they already expected LLMs to help with finding the root cause of failures, while \acs{DeveloperB} focused on the pleasant user experience provided with the integration of \acs{AutoDebug} into Critique for automatic diagnosis upon test failures. Integrating such tools into developers' daily workflows seamlessly and providing them findings automatically, without effort from them to run tools, is typically welcomed by developers.

Participant \acs{DeveloperC} pointed to an \acs{AutoDebug} diagnosis that reported "more information is needed to diagnose the root cause of the failure". Upon investigation, we determined that this was due to one of the infrastructure bugs discussed in Section \ref{section:manual-evaluation}, and there have been around 20 such diagnoses left by \acs{AutoDebug} on various code changes to date. While it helped determine issues in infrastructure, we observe that developers are highly sensitive to unhelpful \acs{AutoDebug} findings, especially when such findings are surfaced automatically and inside their workflow, without any action from the developer, reinforcing the guidelines on limits for unhelpful findings discussed in Section \ref{section:user-feedback}.

Several participants (e.g. \acs{DeveloperD}) discussed that, even though the summaries are helpful, their expectation is receiving a fix automatically. While \acs{AutoDebug} may be perceived as useful, some developers are now expecting more from LLM-based tooling, e.g. fixes instead of diagnoses, as they get a better understanding of the capabilities of LLMs and as they use them more often in their daily workflows.

\begin{figure}
\begin{testimonialbox}
{[\acs{DeveloperE}]}: I don't really use test logs a lot for debugging integration tests, I find it quite hard to find the actual point of failure in those.\\[1em]

{[\acs{DeveloperF}]}: Debugging [integration tests] is definitely a learning curve; there is a large list of log files, I sometimes need to dive deep and jump between them to find the actual error, very tricky. [...] I can imagine it will be very challenging for people who just start integration testing.\\[1em]

{[\acs{DeveloperG}]}: I wish it were clearer where to find the actual error message, often multiple layers, actual error message might be hidden deep down in some log file [...]\\[1em]

{[\acs{DeveloperH}]}: The [diagnosis] information is helpful; I usually have to check logs, but now it's putting all the information in front of me, no need to go into logs, filter by error, no need to scan all of that, I have the summary.\\[1em]

{[\acs{DeveloperA}]}: I always wondered, why couldn't an LLM just have told me about this [error] before? Today's debugging took 10-15 minutes, but we've all had those issues that take over an hour and maybe another engineer's time. [...] This is a great step forward in debugging integration tests faster! I am looking forward to using this tool more going forward.\\[1em]

{[\acs{DeveloperB}]}: [...] It is great that the conclusive debug information shows up in Critique. Now that I have seen it during code review, I reopened my editor to simply make the change.\\[1em]

{[\acs{DeveloperC}]}: \acs{AutoDebug} is telling me that it needs access to more logs to find the root cause. Doesn't it have all the logs?\\[1em]

{[\acs{DeveloperD}]}: \acs{AutoDebug} should not just show a summary, it should give me the fix.
\end{testimonialbox}
\Description{Testimonials}
\caption{Excerpts of feedback on \acs{AutoDebug} Critique findings from interviews conducted with \acs{TotalInterviewedDevelopers} participants.}
\label{fig:testimonials}
\end{figure}

Based on the manual evaluations, user feedback and user interviews, we conclude that LLMs are highly successful in diagnosing integration test failures, given their capacity to process and summarize text. Integrating such tooling to automatically run and help developers in their workflows is perceived positively by developers, although the accuracy of the tool plays a critical role in this perception.
\section{Threats To Validity}

In this section, we discuss the threats to the validity of our work and case studies.
\\

\noindent \textbf{Manual evaluation}: Our manual evaluation was conducted by three expert developers in diagnosing integration tests. However, these developers did not own the production or test code of the failing tests, and may have made mistakes in their reporting of helpfulness of \acs{AutoDebug}.

Additionally, the failures used in this evaluation were randomly chosen from across Google, and may not be representative of all failures.
\\

\noindent \textbf{Developer Bias}: The user feedback on code changes was open to all developers across Google. However, we got feedback from a small subset of developers, which may have introduced a selection bias and biased the accuracy we obtained from user feedback. 
\\
Additionally, the interviews were conducted with code change authors who reported "Useful" or "Not useful" feedback, and reviewers who reported "Please fix". This constitutes a self-selected group, users who did not provide any feedback or who had neutral experiences might have different perspectives, which could alter the qualitative insights gained from the interviews.
\\

\noindent \textbf{Developer infrastructure}: Our case study was conducted within the specific context of Google's development environment and infrastructure, which may limit the generalizability of the findings to other organizations.
\\

\noindent \textbf{LLM}: Our study uses Gemini 2.5 Flash. Although popular large LLMs have been shown to perform well on tasks such as summarization, the generalizability of our results to other LLMs may be limited.

Additionally, LLMs are constantly evolving. The performance and characteristics of Gemini 2.5 Flash may have changed during the course of our studies, which may have impacted our case study results.

Finally, \acs{AutoDebug} relies on a specific prompt template for Gemini. Small changes in the prompt wording, structure, or ordering of instructions could potentially lead to different or degraded LLM performance. The robustness of the prompt engineering to minor variations or future LLM updates is a potential threat to the generalizability of our findings.
\\

\noindent \textbf{Logs}: As they are implemented by developers, the quality and detail of logs from different components may vary significantly, and this may have influenced the accuracy of \acs{AutoDebug} on different tests and the results of our case studies.
\section{Related Work}

There is a rich body of existing work on program diagnosis and repair.
\\

\noindent \textbf{Statistical debugging}: Early automated debugging methods use statistical analysis of program behaviors to pinpoint bugs. For example, Liblit et al. introduce Scalable Statistical Bug Isolation \citep{liblit2005scalable}, which monitors sampled boolean predicates during many runs to find those strongly correlated with failures. Similarly, Liu et al. present SOBER \citep{liu2006statistical}, a hypothesis-testing approach that models each predicate’s distribution in passing vs. failing executions and ranks predicates whose failure-time patterns diverge from normal. These methods assume lightweight instrumentation and rely on correlating events with test outcomes to highlight likely faults. They handle multiple bugs by separating their effects and ranking fault-relevant predicates probabilistically.
\\

\noindent \textbf{Spectrum-based fault localization (SBFL)}: A long line of work uses execution spectra to compute suspiciousness scores for program elements. These techniques rank each statement or entity by combining its execution frequency in failing vs. passing tests. A prominent example is Tarantula \citep{jones2002visualization}, which scores a statement based on the fraction of failing runs relative to all executions. Abreu et al. proposed additional metrics like Ochiai and Jaccard to improve localization \citep{abreu2009practical}; empirical studies show Ochiai often outperforms Tarantula in accuracy. De Souza et al.’s survey \citep{de2016spectrum} highlights that SBFL techniques pinpoint program elements more likely to contain faults using these coverage-based metrics. In recent years, researchers have developed many variants and combinations of SBFL metrics to boost precision.
\\

\noindent \textbf{Delta debugging}: Zeller and Hildebrandt \citep{zeller2002simplifying} propose delta debugging techniques to automatically isolate minimal failure inducing differences in inputs or program changes. Given a failing and a passing execution, delta debugging applies a systematic search to remove irrelevant parts of the input or change set until only the smallest failure-inducing chunk remains. This algorithm repeatedly tests subsets of changes until it finds a minimal failure-inducing scenario. Delta debugging has been applied not only to inputs (e.g., HTML, command-line options) but also to code changes, configurations, and other causal factors.
\\

\noindent \textbf{Failure sketching}: A related research area focuses on summarizing failures through program execution analysis rather than solely relying on logs, originating from thin slicing \citep{sridharan2007thin}, a program analysis technique that generates a minimal set of statements pertinent to a specific value by prioritizing value-flow relevance while excluding indirect dependencies.

Expanding upon this, Kasikci et al. \citep{kasikci2015failure} propose failure sketches, which produce concise root-cause summaries by highlighting only the statements that actually cause a failure and the program-state differences between failing and successful runs.
\\

Compared to traditional diagnosis techniques, our approach is inherently more robust and maintainable, as it can semantically process heterogeneous log formats without the need for component-specific parsing rules and distributed dynamic analysis tools, thus avoiding significant and continuous engineering overhead.
\\

\noindent \textbf{Machine learning for fault localization (FL)}: More recent work applies statistical learning to improve localization. For instance, Sohn and Yoo \citep{sohn2017fluccs} use supervised learning on features derived from past bug fixes and test data to re-rank suspicious code. Other work explored deep-learning and graph-based models for FL \citep{li2021fault, zhang2019cnn}. These methods integrate information like code complexity, textual similarity, and historical bug patterns into neural models. Notably, graph neural networks (GNNs) have been used to capture structural code context: Lou et al. \citep{lou2021boosting} and Qian et al. \citep{qian2023gnet4fl} show GNN-based FL models achieving state-of-the-art localization accuracy by encoding program dependency graphs. Such learning-based approaches can improve coverage-only SBFL rankings, but they often require large training datasets and are primarily evaluated on unit-level faults.
\\

\noindent \textbf{LLM and agent-based debugging}: A recent wave of work leverages LLMs for debugging tasks. Studies find that LLMs trained on code (e.g. GPT-4 \citep{openai2023gpt4}, CodeLlama \citep{roziere2023codellama}) excel at understanding code and suggesting fixes \citep{kang2024quantitative,wu2023large,pu2023summarization,li2023explaining}. For example, Wu et al. \citep{wu2023large} evaluate ChatGPT-3.5/4 \citep{openai2022chatgpt35,openai2023chatgpt4} on fault localization and report that GPT-4 achieves much higher accuracy than prior SBFL tools on Defects4J \citep{just2014defects4j}, especially in small contexts. Kang et al. \citep{kang2024quantitative} introduce AutoFL, an LLM-based FL method that takes a failing test as input and outputs the faulty method plus a natural-language rationale for the fault.
\\

\noindent \textbf{LLM and agent-based repair}: There is a rich body of work on using LLMs and AI agents for program repair, with recent advancements demonstrating better performance than earlier techniques.

Lee et al. present FixAgent \citep{lee2024unified}, which coordinates specialized LLM agents ("Tester" and "Debugger") via prompt chaining to perform end-to-end fault localization and repair. Similarly, LLM4FL \citep{nakhla2024enhancing} splits large coverage data into manageable chunks and uses two LLM agents to iteratively analyze code to rank suspicious methods. Yang et al. \citep{yang2024swe} proposed SWE-Agent, an AI agent to solve bugs in SWE-Bench \citep{jimenez2023swe} and HumanEvalFix \citep{muennighoff2023octopack} benchmarks. RepairAgent \citep{bouzenia2024repairagent} uses a state machine to control its agent's actions, limiting certain operations. AutoCodeRover \citep{zhang2024autocoderover} leverages explicit program information like class and method definitions, alongside test-based localization. Extending this, SpecRover \citep{ruan2024specrover} introduces a natural-language specification for desired behavior at each potential repair site and includes a patch review agent. CodeR \citep{chen2024coder} breaks down the task of program repair into various sub-tasks, managed and reviewed by a "manager" agent through a task graph. MarsCode Agent \citep{liu2024marscode} integrates a dynamic, iterative repair process with a traditional generate-and-validate approach within a multi-agent framework.  AlphaRepair \citep{xia2022less} offers a zero-shot solution, enabling LLMs to rectify bugs without the need for supplementary training data. Building upon this, FitRepair \citep{xia2023automated} enhances AlphaRepair by integrating the plastic surgery hypothesis \citep{barr2014plastic}, which posits that the necessary bug-fixing code often resides within the existing project. Another approach, ChatRepair \citep{xia2024automated}, facilitates error correction through an interactive dialogue with the engineer.

In contrast to these methods, Agentless \citep{xia2024agentless} employs a streamlined, agentless strategy, bypassing intricate tools and decision-making for a three-stage process: localization, repair, and patch validation. Agentless has achieved state-of-the-art results with notably reduced costs, surpassing the performance of current open-source software agents.

Rondon et al. propose Passerine \citep{rondon2025evaluating} to fix bugs at Google using LLM agents. Maddila et al. discuss Engineering Agent \citep{maddila2025agentic} to automatically fix failing unit-tests at Meta.
\\

\noindent \textbf{Benchmarks for debugging and repair}: To systematically assess LLM debugging and repair, new benchmarks have emerged. SWE-Bench \citep{jimenez2023swe} is a benchmark sourced from GitHub issues across 12 popular open source projects containing Python bugs and fixes. HumanEvalFix \citep{muennighoff2023octopack} is a short-form code debugging benchmark. DebugEval \citep{yang2024enhancing} defines multiple tasks (bug localization, bug classification, code review, code repair) to evaluate LLM debugging ability across languages. DebugBench \citep{majdoub2024debugging} is a large dataset of LeetCode coding problems with injected bugs to compare several open-source LLMs on automated bug fixing. SWE-smith \citep{yang2025swe} is a pipeline for generating software engineering tasks at scale, allowing the automatic synthesis of changes that break existing tests in a codebase. SWE-Lancer \citep{miserendino2025swe} is a benchmark that consists of over 1,400 freelance software engineering tasks. Almost all of these benchmarks focus on unit-test related debugging and repair tasks. 
\\

Previous studies discussed above generally show that large models (70B+ parameters) have convincing localization, debugging and repair capabilities. Overall, LLM-based techniques currently outperform many traditional methods for unit-level bug localization and repair, but they face new challenges (e.g. limited token context, integration of dynamic traces) when scaling to larger systems \citep{hadi2023large, hou2024large, wu2023large}.

Most of the above work, both benchmarks and research, concentrate on small-scale or unit-test debugging and repair scenarios. Integration tests, where faults emerge from interactions among components, pose additional challenges, e.g. complex environment setup, non-deterministic dependencies, significant log output, and more complex failure modes that remain under-explored.

Our approach builds on these foundations but focuses on leveraging LLMs for the difficult problem of diagnosing integration test failures, aiming to address scenarios that traditional and even recent LLM-based techniques have not fully tackled.

\section{Conclusion}

Integration testing is a fundamental phase in ensuring the quality and reliability of complex, distributed software systems, yet diagnosing their failures presents significant challenges for developers. Developers consistently report spending substantially more time on diagnosing integration test failures compared to unit test failures.

To overcome these, this paper introduced \acs{AutoDebug}, a novel failure diagnosis tool powered by LLMs. The system determines the root cause of integration test failures, produces concise summaries and highlights the specific log lines most relevant to the diagnosis. It is integrated directly into developer code review workflows, providing contextual and in-time assistance to streamline the debugging process.

\acs{AutoDebug}'s effectiveness is evaluated through quantitative and qualitative methods. A manual evaluation on \acs{TotalEvalExamples} real-world integration test failures demonstrated that the tool achieved a \acs{TotalEvalAccuracy} accuracy in detecting and summarizing the root cause diagnosis. This evaluation also proved beneficial in uncovering underlying infrastructure bugs related to log saving.

Following its launch, \acs{AutoDebug} processed a large number of test failures. User feedback indicated a "Not helpful" rate of \acs{TotalFeedbacksNotUsefulPct}, which is well within the acceptable guideline for tools posting findings in Critique. Although the reported helpfulness-rate was \acs{TotalFeedbacksHelpfulnessRate}, this disparity compared to the manual evaluation's \acs{TotalEvalAccuracy} accuracy is attributed to the well-established negativity bias, where users are more inclined to report negative experiences. \acs{AutoDebug}'s high ranking of \acs{AutoDebugHelpfulnessRank} (top \acs{AutoDebugHelpfulnessRankPct}) among \acs{TotalAnalyzers} tools in Critique further suggests that its diagnoses resonate with developers who have historically been frustrated with integration test failure diagnosis.

User interviews confirmed \acs{AutoDebug}'s perceived usefulness and the positive reception of its integration into existing workflows. Developers appreciated the automatic, in-context diagnostic assistance, with some even beginning to expect more advanced capabilities, such as automated fixes, from LLM-based tools.

In conclusion, LLMs are highly successful in diagnosing integration test failures, primarily due to their capacity to process and summarize complex textual data. The seamless integration of such AI-powered tooling into developers' daily workflows is perceived positively, although the accuracy of the tool remains a critical factor in shaping developer perception and adoption.
\section{Future Work}

We observed two major improvement opportunities during our case studies of \acs{AutoDebug}.

First, in failing integration test logs, several log lines are output at the \texttt{ERROR} level even though the reported issue is later recovered during the test run. Such logs are not only confusing for developers, but also for LLMs. We plan to filter such logs by using the logs from a previous passing execution of the test, and dropping a log if a similar log exists in a passing run. We predict that this will improve the accuracy of the diagnoses the LLM produces.

Second, we plan to extend \acs{AutoDebug} to provide not only diagnoses but also fixes when possible, aligned with developer expectations.
\begin{acks}

We thank Elvira Djuraeva for her help on the user surveys to understand the unit and integration test usage across Google, James McClure and Eric Beerman on their help with the manual evaluations, and Jessica Tan, Maria Arguello and Elaine Thai for their help on conducting user interviews.

Google Gemini \citep{googleGemini2025} was utilized to generate sections of this work, specifically several tables in Latex \citep{lamport1994latex} format, and several graphs inside Google Colab \citep{googleColab}.

\end{acks}

\bibliographystyle{ACM-Reference-Format}
\bibliography{0-paper}
\end{document}